\documentclass[prd,aps,showpacs,nofootinbib,eqsecnum,onecolumn,groupedaddress,amssymb]{revtex4}
\usepackage{amsmath}
\usepackage{amssymb}
\usepackage{amsfonts}
\usepackage{graphicx,bm}
\usepackage{dcolumn}
\usepackage{color,amsxtra}
\usepackage{epsf}
\usepackage{enumerate}
\usepackage{hhline}
\usepackage{array}
\usepackage{tabularx}
\usepackage{subfigure}
\usepackage{fancyhdr}
\usepackage{mathrsfs}
\newcommand{\be}{\begin{equation}}
\newcommand{\ee}{\end{equation}}
\newcommand{\bea}{\begin{eqnarray}}
\newcommand{\eea}{\end{eqnarray}}
\newcommand{\beaa}{\begin{eqnarray*}}
\newcommand{\eeaa}{\end{eqnarray*}}

\def\be{\begin{equation}}
\def\ee{\end{equation}}
\def\bea{\begin{eqnarray}}
\def\eea{\end{eqnarray}}

\begin{document}

\title{Inflation in $f(R,\phi)$-theories and mimetic gravity scenario}
\author{
R.~Myrzakulov$^{1}$
\footnote{E-mail address: myrzakulov@gmail.com},
L.~Sebastiani$^{1}$
\footnote{E-mail address: l.sebastiani@science.unitn.it}
and S.~Vagnozzi$^{2,3,4,5}$
\footnote{E-mail address: vagnozzi@nbi.dk}
}
\affiliation{
$^1$Department of General \& Theoretical Physics and Eurasian Center for
Theoretical Physics, Eurasian National University, Astana 010008, Kazakhstan\\
$^2$ Niels Bohr International Academy and Discovery Center, Niels Bohr Institute, University of Copenhagen, Blegdamsvej 17, 2100 Copenhagen \O, Denmark \\
$^3$ Nordita,
KTH Royal Institute of Technology and Stockholm University,
Roslagstullsbacken 23, SE-106 91 Stockholm, Sweden \\
$^4$ The Oskar Klein Centre for Cosmoparticle Physics, Department of Physics, Stockholm University, AlbaNova, SE-106 91 Stockholm, Sweden \\
$^5$ ARC Centre of Excellence for Particle Physics at the Terascale, School of Physics, University of Melbourne, Victoria 3010, Australia}

%\date{}

%%%%%%%%%%%%%%%%%%%%%
% Abstract
%%%%%%%%%%%%%%%%%%%%%
\begin{abstract}
We investigate inflation within $f(R,\phi)$-theories, where a dynamical scalar field is coupled to gravity. A class of models which can support early-time acceleration with the emerging of an effective cosmological constant at high curvature is studied. The dynamics of the field allow for exit from inflation leading to the correct amount of inflation in agreement with cosmological data. Furthermore, the spectral index and tensor-to-scalar ratio of the models are carefully analyzed. A generalization of the theory to incorporate dark matter in the context of mimetic gravity, and further extensions of the latter, are also discussed.
\end{abstract}
%%%%%%%%%%%%%%%%%%%%%

%----------------------------
%\pacs{98.80.Cq, 12.60.-i, 04.50.Kd, 95.36.+x}
%\hspace{13.1cm}
%----------------------------

\maketitle

\def\thesection{\Roman{section}}
\def\theequation{\Roman{section}.\arabic{equation}}

%%%%%%%%%%%%%%%%%%%%%%%%%%%
%%%  Sec. I
%%%%%%%%%%%%%%%%%%%%%%%%%%%
\section{Introduction}

Over the past years, interest towards inflationary cosmology has grown considerably, as a consequence of the great amount of data from recent cosmological surveys ~\cite{WMAP, Planck, Planck2}. The inflationary paradigm was first introduced in 1981 by Guth~\cite{Guth} and Sato~\cite{Sato} to explain the thermalization of the observable Universe inferred from observations of the CMB. It also allows to address some of the problems associated to the initial conditions of a Friedmann universe. Moreover, quantum fluctuations during the inflationary epoch presumably seeded the perturbations which grew under gravitational instability into the structures we see today \cite{mukhanov}. For reviews on inflation, see e.g.~\cite{book, Linde, revinflazione, tasi, kinney}. \\

An early-time period of acceleration should presumably be supported by a repulsive gravitational force. At the same time, a mechanism which allows to quickly exit this stage and enter the radiation dominated era is necessary.  The arena of inflationary models is quite vast. In the scalar field formulation or chaotic inflation~\cite{chaotic}, a scalar field (the inflaton) is subject to a potential and drives accelerated expansion when its magnitude is negative and very large: at the end of inflation, it settles down in a minimum of the potential and begins oscillating, giving rise to the reheating mechanism responsible for particle production. Other implementations of inflation include for instance natural inflation (see e.g. \cite{freese}), k-inflation \cite{kmuk}, brane inflation \cite{gdva}, and many others\footnote{Very recently, inflation has been implemented by means of a BIon system, see \cite{bion}.}. In the context of modified gravity (see Refs.~\cite{revmod} for reviews), a modification to Einstein's gravitational action emerges at high curvature and supports the early-time acceleration (see Ref.~\cite{OdOd}). This can be realized for instance in the so-called Starobinsky model~\cite{Staro}, which provides a correction quadratic in the Ricci scalar. \\

A model for inflation is viable only if it is able to reproduce the inferred spectral index and the tensor-to-scalar ratio at the origin of cosmological perturbations in the Friedmann universe. The evaluation of these indices depends on the theory under investigation. In ~\cite{corea} an unified description has been derived in the context of $f(R,\phi)$-gravity, where a scalar field subject to a potential is coupled to gravity (a nonminimal coupling in the kinetic part of the field is also present). In this work, we will analyze $f(R,\phi)$-inflation by working through some simple examples based on modified gravity models which mimic the ``false vacuum'' of the primordial universe: in fact, we will study a class of models (exponential models and power-law models) describing an effective cosmological constant at high curvature and whose exit from inflation is induced by the coupling of a scalar field to gravity. We will show that the model can yield values for the spectral index and the tensor-to-scalar ratio (the magnitude of the last one in $f(R,\phi)$-gravity is particularly small) in agreement with those inferred from observations. \\

In the last part of the work we embed this model within the framework of mimetic gravity, which additionally endows it with a dark matter candidate. We then discuss extensions which can address the controversies of cold dark matter on small scales within the mimetic gravity scenario, such extensions being theoretically driven by the correspondence between mimetic gravity and the scalar formulation of the Einstein-aether theories. We then speculate on possible further extensions of the scenario depicted. \\

%%%%%
The paper is organized as follows. In Section {\bf II} we will revisit the form of the spectral indices and the tensor-to-scalar ratio in $f(R,\phi)$-gravity by deriving some useful relations. In Section {\bf III} we study inflation in two different $f(R,\phi)$-models. Early-time acceleration takes place at high curvature in agreement with the latest Planck data and the field allows for a quick exit from this stage recovering Friedmann evolution of Einstein's gravity. In Section {\bf IV} we formulate the model within the mimetic gravity framework. Section {\bf V} is devoted to conclusions and final remarks. \\

%%% Unit %%%
We use units where $k_{\mathrm{B}} = c = \hbar = 1$ and denote the
gravitational constant, $G_N$, by $\kappa^2\equiv 8 \pi G_{N}$, such that
$G_{N}=M_{\mathrm{Pl}}^{-2}$, $M_{\mathrm{Pl}} =1.2 \times 10^{19}$ GeV being the Planck mass.
%%%%%%%%%%%%
%%%%%

%%%%%%%%%%%%%%%%%%%%%%%%%%%
%%%  Sec. II
%%%%%%%%%%%%%%%%%%%%%%%%%%%
\section{Inflation in $f(R, \phi)$-gravitational models }

Let us consider the following Lagrangian for a scalar field coupled to gravity,
\begin{equation}
\mathcal L= \frac{f(R,\phi)}{2}-\frac{\omega(\phi)\partial^{\mu}\phi\partial_{\mu}\phi}{2}-V(\phi)\,,\label{L}
\end{equation}
where $f(R,\phi)$ is a generic function depending on the Ricci scalar $R$ and the scalar field $\phi$ is subject to the potential $V(\phi)$. In the above, $\omega(\phi)$ is in principle a function of $\phi$ which represents a non-minimal coupling of the kinetic term of the field. In a flat Friedmann-Robertson-Walker universe, with metric given by
\begin{equation}
ds^2=-dt^2+a(t)^2 {\bf dx}^2\,,\label{metric}
\end{equation}
where $a\equiv a(t)$ is the scale factor depending on the cosmological time $t$,
the Equations of Motion of the theory read:
\begin{equation}
3 F(R,\phi) H^2=\frac{\omega(\phi)\dot\phi^2}{2}+V(\phi)+\frac{1}{2}\left(R F(R,\phi)-f(R,\phi)\right)-3 H\dot F(R,\phi)\,,\label{EOM1}
\end{equation}
\begin{equation}
-2 F(R,\phi)\dot H=\omega(\phi)\dot\phi^2+\ddot F(R,\phi)-H \dot F(R,\phi)\,.\label{EOM2}
\end{equation}
Here, $H=\dot a/a$ is the Hubble parameter, the dot denoting the time derivative. In the above, we have made use of the following definition,
\begin{equation}
F(R,\phi)=\frac{d}{d R}f(R,\phi)\,.\label{Fprime}
\end{equation}
From (\ref{EOM1})--(\ref{EOM2}) we derive the continuity equation of the field, which reads
\begin{equation}
\ddot\phi+3H\dot\phi+\frac{1}{2\omega(\phi)}\left(\dot\omega(\phi)\dot\phi-\frac{d f(R,\phi)}{d\phi}+2\frac{d V(\phi)}{d\phi}\right)=0\,.\label{cons}
\end{equation}
Inflation is described by a (quasi) de Sitter solution where the Hubble parameter is nearly constant. To proceed, one introduces the ``slow-roll'' parameters~\cite{sr},
\begin{equation}
\epsilon_1=-\frac{\dot H}{H^2}\,,\quad\epsilon_2=\frac{\ddot\phi}{H\dot\phi}\,,\quad
\epsilon_3=\frac{\dot F(R,\phi)}{2 H F(R,\phi)}\,,\quad
\epsilon_4=\frac{\dot E}{2 H E}\,,\label{srpar}
\end{equation}
where
\begin{equation}
E=F(R,\phi)\omega(\phi)+\frac{3\dot F(R,\phi)^2}{2\dot\phi^2}\,.
\end{equation}
During inflation, the magnitude of the slow roll parameters is presumed to be very small (in what is known as the slow-roll approximation). In particular, given that
\begin{equation}
\frac{\ddot a}{a}=\dot H+H^2\,,
\end{equation}
one needs $0<\epsilon_1\ll 1$ in order to have a strong accelerated expansion with $\dot H <0$. Correspondingly, the inflation epoch and the acceleration end at a time for which $\epsilon_1 \simeq 1$. We observe that
\begin{equation}
\epsilon_4=
\frac{\left[\frac{\dot\phi^2}{H^2\dot F(R,\phi)}\left(\dot\omega(\phi)-4H\omega(\phi)\epsilon_3\right)
+6\epsilon_1+6\epsilon_3(1-\epsilon_2)
\right]}{2\left[\frac{\omega(\phi)\dot\phi^2}{H\dot F(R,\phi)}+3\epsilon_3\right]}\,,
\end{equation}
where
\begin{equation}
\dot\omega(\phi)=-2H\omega(\phi)\left(3+\epsilon_2\right)+\frac{1}{\dot\phi}\left[\frac{df(R,\phi)}{d\phi}-2\frac{d V(\phi)}{d\phi}\right]\,.
\end{equation}
Thus, under the slow roll approximation we have that $|\dot\omega(\phi)\dot\phi^2/(H^2\dot F(R,\phi))|\ll 1$ and, since $|\omega(\phi)\dot\phi^2/F(R,\phi)H^2|\equiv |2\omega(\phi)\dot\phi^2\epsilon_3/(H \dot F(R,\phi))|\ll |\dot\omega(\phi)\dot\phi^2/(H^2\dot F(R,\phi))|$, combining Eq. (\ref{EOM1}) with Eq. (\ref{cons}) yields\footnote{If $\dot\omega(\phi)=0$, we obtain directly 
$|\omega(\phi)\dot\phi^2/F(R,\phi)H^2|\ll 1$ from $|\epsilon_4|\ll 1$.}
\begin{equation}
3F(R,\phi) H^2\simeq V(\phi)+\frac{1}{2}\left(R F(R,\phi)-f(R,\phi)\right)\,,\quad3H\dot\phi+\frac{1}{2\omega(\phi)}\left(\dot\omega(\phi)\dot\phi-\frac{d f(R,\phi)}{d\phi}+2\frac{d V(\phi)}{d\phi}\right)\simeq 0\,.\label{EOMsr}
\end{equation}
As a measure of perturbations during inflation, one introduces the spectral index $n_s$ and the tensor-to-scalar ratio $r$ defined as~\cite{corea},
\begin{equation}
n_s=1-4\epsilon_1-2\epsilon_2+2\epsilon_3-2\epsilon_4\,,\quad
r=16(\epsilon_1+\epsilon_3)\,,
\end{equation}
where $\epsilon_{1,2,3,4}$ must be evaluated during inflation, in the slow roll regime. \\

In the simple scalar field theory with $F'(R,\phi)=1/\kappa^2$ and $\omega(\phi)=1$, one has
\begin{equation}
\epsilon_1=\frac{1}{2\kappa^2 V(\phi)^2}\left(\frac{d V(\phi)}{d\phi}\right)^2\,,\quad\epsilon_2=\epsilon_1-\frac{1}{\kappa^2 V(\phi)}\left(\frac{d^2 V(\phi)}{d\phi^2}\right)\,,
\quad \epsilon_2=\epsilon_4=0\,.
\end{equation}
Hence, we recover
\begin{equation}
n_s=1-6\epsilon+2\eta\,,\quad r=16\epsilon
\end{equation}
where $\epsilon\,,\eta$ are given by
\begin{equation}
\epsilon\equiv\epsilon_1=\frac{1}{2\kappa^2 V(\phi)^2}\left(\frac{d V(\phi)}{d\phi}\right)^2\,,\quad \eta=\frac{1}{\kappa^2 V(\phi)}\left(\frac{d^2 V(\phi)}{d\phi^2}\right)\,.
\end{equation}
In the modified gravity case instead, with $F(R,\phi)\equiv F(R)$ and $\omega(\phi)=V(\phi)=0$, given that $\epsilon_2=0$, $\epsilon_1\simeq -\epsilon_3(1-\epsilon_4)$ and, in the slow roll approximation, $\epsilon_1\simeq-\epsilon_3$ and $\epsilon_4\simeq-3\epsilon_1+\dot\epsilon_1/(H\epsilon_1)$, we obtain
\begin{equation}
n_s\simeq 1-6\epsilon_1-2\epsilon_4=1-\frac{2\dot\epsilon_1}{H\epsilon_1}\,,\quad
r=48\epsilon_1^2\,,
\end{equation}
where for the expression of the tensor-to-scalar ratio one has to use the higher order corrections of the slow-roll parameters.
Finally, if $\dot\omega(\phi)=0$ (non-coupling of the field with kinetic energy) and $\omega(\phi)\equiv\omega$, the following relation holds true
\begin{equation}
\epsilon_1=-\epsilon_3+\frac{\omega\dot\phi^2}{3H\dot F(R,\phi)}\left(\epsilon_4+2\epsilon_3\right)+\epsilon_3(\epsilon_2+\epsilon_4)\simeq
-\epsilon_3+\frac{\omega\dot\phi^2}{3H\dot F(R,\phi)}\left(\epsilon_4+2\epsilon_3\right)\,,
\end{equation}
where we have taken into account the slow roll approximation\footnote{
If $\dot\omega(\phi)\neq 0$, one has $\epsilon_1\simeq-\epsilon_3-\dot\omega(\phi)\dot\phi^2/(6H^2\dot F(R,\phi))$.}. As a consequence, the spectral index and tensor-to-scalar ratio read
\begin{eqnarray}
n_s&\simeq& 1-2\epsilon_1\left(\frac{3H\dot F(R,\phi)+2\omega\dot\phi^2}{\omega\dot\phi^2}\right)-2\epsilon_2-6\epsilon_3\left(\frac{H\dot F(R,\phi)-\omega\dot\phi^2}{\omega\dot\phi^2}\right)\nonumber\\
&=&1+
\frac{2\dot H}{H^2}\left(\frac{3H\dot F(R,\phi)+2\omega\dot\phi^2}{\omega\dot\phi^2}\right)
-\frac{\ddot\phi}{H\dot\phi}-\frac{3\dot F(R,\phi)}{H F(R,\phi)}\left(\frac{H\dot F(R,\phi)-\omega\dot\phi^2}{\omega\dot\phi^2}\right)\,,\label{index2}
\\
r&=&16(\epsilon_1+\epsilon_3)=-\frac{16\dot H}{H^2}+\frac{8\dot F(R,\phi)}{H F(R,\phi)}\,.\label{ratio2}
\end{eqnarray}
The latest cosmological data from Planck satellite~\cite{Planck2} constrain these two quantities as
$n_{\mathrm{s}} = 0.968 \pm 0.006\, (68\%\,\mathrm{CL})$ and 
$r < 0.11\, (95\%\,\mathrm{CL})$.

\section{Viable inflation in $f(R,\phi)$-gravity describing an effective cosmological constant}

In order to reproduce the ``false vacuum'' of inflation, one possibility is to introduce a large effective cosmological constant (whose value is close to the Planck scale) within Einstein's framework. In this way, it is easy to obtain the repulsive gravity required to support the early-time acceleration. However, one of the main problems faced by the inflationary paradigm is the realization of a mechanism to gracefully exit this stage. \\

Over the past years, a class of viable exponential models of $f(R)$-modified gravity which can succesfully realize the current acceleration of our universe have been investigated. These models feature what can be viewed as a ``switching on'' cosmological constant and assume the following form~\cite{group1, group2, group3}
\begin{equation}
f(R)=\frac{R}{\kappa^2}-\frac{2\Lambda}{\kappa^2}\left(1-\text{e}^{-\frac{R}{R_0}}\right)\,,\label{main}
\end{equation}
where $\Lambda$ is the cosmological constant and $R_0$ the curvature scale at which such a constant is expected to appear. It is easy to see that for $R_0\ll R$ the model behaves as one where $f(R)\simeq R-2\Lambda$. Furthermore, by setting $\Lambda/\kappa^2$ to being the current amount of dark energy in our universe, one recovers the $\Lambda$CDM model. In particular, the dark energy epoch is realized by a stable de Sitter solution, with the model passing all cosmological tests. \\

It is our intention to extend the model examined in \ref{main} to include inflation. In order to exit from the early-time acceleration period (namely, in order to ``swith off'' the cosmological constant), we will introduce a dynamical field $\phi$ by making the following substitution,
\begin{equation}
\frac{1}{R_0}\rightarrow -b\kappa^3\phi\,. \label{sub}
\end{equation}
Here, $\phi$ is assumed to be negative and dependent on the cosmological time, $\kappa^3$ has been introduced for dimensional reasons and $b$ is a dimensionless number of order unity. In this way,
the scale at which the cosmological constant appears is a sort of ``running scale'', which varies as the field does. We consider the Lagrangian
\begin{equation}
\mathcal L=\frac{R-2\Lambda\left(1-\text{e}^{b\phi\kappa^3 R}\right)}{2\kappa^2}+\frac{\dot\phi^2}{2}\,, 
\end{equation}
which corresponds to (\ref{L}) with
\begin{equation}
f(R,\phi)=\frac{R-2\Lambda\left(1-\text{e}^{b\phi\kappa^3 R}\right)}{\kappa^2}\,,\quad
\omega(\phi)=1\,,\quad
V(\phi)=0\,.
\label{model1}
\end{equation}
At the onset of inflation, the field $\phi=\phi_{0}$ is negative and very large,
\begin{equation}
\frac{1}{b\kappa^3 R}\ll \phi_0\,.\label{cond0}
\end{equation}
Given that the curvature during inflation is close to the Planck scale, one may argue that in general
\begin{equation}
M_{Pl}<-\phi_0\,.
\end{equation}
The fact that the field is super-Planckian does not pose a problem, provided it is understood that its kinetic energy be $\dot\phi^2/2<M_{Pl}^2$ in order to avoid quantum effects in the theory. \\

Under the condition set by (\ref{cond0}) and by assuming the slow roll approximation $|\epsilon_{1,2,3,4}|\ll 1$ in (\ref{srpar}), (\ref{EOMsr}) allows one to get,
\begin{equation}
H\simeq\sqrt{\frac{\Lambda}{3}}\,,\quad \dot\phi\simeq 4\Lambda H b\kappa \text{e}^{12 H^2 b\phi\kappa^3}\,,\label{sol0}
\end{equation}
where we have used the fact that $R=12H^2+\dot H\simeq 12 H^2$ and $R=4\Lambda$ becomes the curvature scale at which inflation appears. We observe that the field changes faster than the Hubble parameter, as can be noted from (\ref{EOM2}), where $\dot H/\kappa^2\sim\dot\phi^2$. Thus, inflation ends when \begin{equation}
\phi<\frac{1}{4\Lambda b\kappa^3}\,,\quad f(R,\phi)\simeq \frac{R}{\kappa^2}\,,
\end{equation}
and one recovers the Friedmann universe with Einstein's gravity.
In what follows, we will take into account the following relation:
\begin{equation}
|\dot H\phi/(\dot\phi H)|\sim 12 H^2 b\phi\kappa^3 \exp\left[12H^2 b\phi\kappa^3\right]\ll 1\,.\label{cond1}
\end{equation}
Thus, under the slow roll approximation and by using condition (\ref{cond0}) and (\ref{cond1}) we get
\begin{equation}
\dot F(R,\phi)\simeq 24\Lambda H^2 b^2\kappa^4\phi\dot\phi\text{e}^{12 H^2 b\phi\kappa^3 }\,,\quad
\ddot F(R,\phi)\simeq
24\Lambda H^2 b^2\kappa^4\left[\phi\ddot\phi+12 b\phi\kappa^3 H^2\dot\phi^2\right]
\text{e}^{12 H^2 b\phi\kappa^3 }\,,\label{FFF}
\end{equation}
where
\begin{equation}
\ddot\phi\simeq 48\Lambda H^3 b^2\kappa^4 \dot\phi
\text{e}^{12 H^2 b\phi\kappa^3 }\,.\label{FF}
\end{equation}
By combining the expressions in (\ref{FFF}) with (\ref{sol0}) and (\ref{FF}) we obtain
\begin{equation}
\dot F(R,\phi)\simeq 6H b\kappa^3\phi\dot\phi^2\,,\quad
\ddot F(R,\phi)\simeq 144 H^3 b^2\kappa^6\phi\dot\phi^3\,,\quad
\ddot\phi\simeq12H^2 b\kappa^3\dot\phi^2\,.
\end{equation}
Thus, we find that the slow roll paramters in (\ref{srpar}) are given by:
\begin{equation}
\epsilon_1\simeq-4\Lambda^2 b^2\kappa^4\left(4\Lambda b \kappa^3\phi_0\right)
\text{e}^{8 \Lambda b\phi_0\kappa^3}\,,\quad
\epsilon_2\simeq
16\Lambda^2 b^2\kappa^4 
\text{e}^{4\Lambda b\phi_0\kappa^3}\,,\quad
\epsilon_3\simeq
16\Lambda^3 b^3\kappa^7\phi_0\text{e}^{8\Lambda b\phi_0\kappa^3}\,,
\end{equation}
where we have used (\ref{EOM2}) to evaluate $\dot H$, and have set $R=4\Lambda$ and $\phi=\phi_0$. We also obtain
\begin{equation}
\frac{H\dot F(R,\phi)}{\dot\phi^2}\simeq \frac{4\Lambda b\kappa^3\phi_0}{2}\,.
\end{equation}
As a result, the spectral index and the tensor-to-scalar ratio in (\ref{index2})--(\ref{ratio2}) with (\ref{cond1}) read
\begin{equation}
n_s\simeq 1-2\epsilon_2=1-32\Lambda^2 b^2\kappa^4 
\text{e}^{4\Lambda b\phi_0\kappa^3}\,,\quad
r\simeq
\mathcal O(\epsilon_{1,3}^2)\,.
\end{equation}
An important parameter which measures the amount of inflation is the number of $e$-folds $\mathcal N$, which is defined to be:
\begin{equation}
\mathcal N\equiv\ln \left(\frac{a_\mathrm{f}(t_\text{f})}{a_\mathrm{i} (t_\text{i})}\right)=\int^{t_\text{f}}_{t_\text{i}} H(t)
dt\,,\label{Nfolds}
\end{equation}
where $a_\text{i}(t_{i})$ and $a_{\text{f}}(t_{\text{f}})$ are the scale factor at the onset and at the end of inflation, respectively, and $t_\text{i,f}$ are the respective times. According to the latest data, the number of $e$-folds must be $55<\mathcal N<65$ in order for the observable universe to thermalize. Accounting for the fact that the Ricci scalar changes considerably slower than the field itself, we can write in our case that
\begin{equation}
\mathcal N=\int^{\phi_\text{f}}_{\phi_0}\frac{H}{\dot\phi}d\phi\simeq
\int^{\phi_\text{f}}_{\phi_0}\frac{\text{e}^{-4\Lambda b\kappa^3 \phi}}{4\Lambda b\kappa }d\phi
\simeq\frac{\text{e}^{-4\Lambda b\phi_0\kappa^3}}{16\Lambda^2b^2\kappa^4}\,.
\end{equation}
Here, $\phi_\text{i}$ is the value of the field at the end of inflation such that $|\phi_\text{i}|\ll |\phi_0|$.
Thus,
\begin{equation}
1-n_s=\frac{2}{\mathcal N}\,,\quad
r\sim\frac{1}{\mathcal N^4}\,,
\end{equation}
in accordance with the latest Planck data. Since the tensor-to-scalar ratio $r$ is required to be very small but not vanishing, by taking into account that the most likely value for $r$ is $r\sim 0.06$, we require that $\Lambda^2\sim 10^{-6}\times\phi_0^2/\kappa^2$.\\
\\
The same analysis may be applied to another well known $f(R)$ model which mimics the behaviour of (\ref{main}), namely the Hu-Sawicki model~\cite{Hu} (see \cite{dunsby} for recent studies of cosmological dynamics within this model),
\begin{equation}
f(R)=\frac{R}{\kappa^2}-\frac{2\Lambda}{\kappa^2}\left[1-\frac{1}{1+\left(R/R_0\right)^n}\right]\,,\quad 0<n\,.\label{HS}
\end{equation}
Here, $n$ is a positive fixed parameter and
the cosmological constant emerges when $1\ll (R/R_0)^n$. To exit from the accelerated phase, one can make the substitution (\ref{sub}) in the above expression to obtain
\begin{equation}
f(R,\phi)=\frac{R}{\kappa^2}-\frac{2\Lambda}{\kappa^2}\left[1-\frac{1}{1+(-b \phi\kappa^3 R)^n}\right]\,,
\label{model2}
\end{equation}
with the Lagrangian given by:
\begin{equation}
\mathcal L=
\frac{R}{2\kappa^2}-\frac{\Lambda}{\kappa^2}\left[1-\frac{1}{1+(-b \phi\kappa^3 R)^n}\right]+\frac{\dot\phi^2}{2}\,.
\end{equation}
Inflation is realized under the condition given by (\ref{cond0}), namely that the field be negative and its magnitude be larger than the Planck scale.
Under the slow roll approximation, Eqs.~(\ref{EOMsr}) can be solved by
\begin{equation}
H\simeq\sqrt{\frac{\Lambda}{3}}\,,\quad \dot\phi\simeq \frac{4\Lambda H b\kappa n}{(-12H^2 b\phi\kappa^3)^{n+1}}\,.
\end{equation}
We find
\begin{eqnarray}
\dot F(R,\phi)&\simeq&
-\frac{2\Lambda b\dot\phi\kappa n}{\left(-12H^2 b\phi\kappa^3\right)^{n+1}}
\equiv -\frac{\dot\phi^2}{2H}\,,\nonumber\\
\ddot F(R,\phi)&\simeq&-\frac{2\Lambda b\kappa n}{\left(-12H^2 b\phi\kappa^3\right)^{n+1}}\left(
\ddot\phi+\frac{12(n+1)H^2 b\dot\phi^2\kappa^3}{\left(-12H^2 b\phi\kappa^3\right)}
\right)
\equiv -\frac{12(n+1) H b\dot\phi^3\kappa^3}{\left(-12H^2 b\phi\kappa^3\right)}\,,
\end{eqnarray}
where we considered the fact that the field varies faster than the Ricci scalar and have made us of
\begin{equation}
\ddot\phi=\frac{48 n (n+1)\Lambda H^3 b^2\kappa^4\dot\phi}{\left(-12H^2 b\phi\kappa^3\right)^{n+2}}\equiv
\frac{12(n+1) H^2 b\dot\phi^2\kappa^3}{\left(-12H^2 b\phi\kappa^3\right)}\,.
\end{equation}
Thus, the slow roll parameters in (\ref{srpar}) read
\begin{equation}
\epsilon_1\simeq
\frac{12 n^2\Lambda^2 b^2\kappa^4 }{\left(-4\Lambda b\phi_0\kappa^3\right)^{2(n+1)}}
\,,\quad
\epsilon_2\simeq
\frac{16n(n+1)\Lambda^2 b^2\kappa^4}{\left(-4\Lambda b\phi_0\kappa^3\right)^{n+2}}
\,,\quad
\epsilon_3\simeq
-\frac{4\Lambda^2 n^2  b^2\kappa^4}{(-4\Lambda b\phi_0\kappa^3)^{2(n+1)}}
\,,
\end{equation}
where $\phi_0$ is, as usual, the value of the field at the onset of inflation. We also have
\begin{equation}
\frac{H\dot F(R,\phi)}{\dot\phi^2}=-\frac{1}{2}\,.
\end{equation}
The spectral index 
and the tensor-to-scalar ratio in (\ref{index2})--(\ref{ratio2}) are found to be
\begin{equation}
n_s\simeq 1-2\epsilon_2\simeq
1-
\frac{32n(n+1)\Lambda^2 b^2\kappa^4}{\left(-4\Lambda b\phi_0\kappa^3\right)^{n+2}}\,,\quad
r=\frac{32\left(4\Lambda^2 n^2  b^2\kappa^4\right)}{(-4\Lambda b\phi_0\kappa^3)^{2(n+1)}}\,.
\end{equation}
Moreover, the number of $e$-folds can be written as:
\begin{equation}
\mathcal N=\int^{\phi_\text{f}}_{\phi_0}\frac{H}{\dot\phi}d\phi\simeq
\int^{\phi_\text{f}}_{\phi_0}\frac{(-4\Lambda b\phi\kappa^3)^{n+1}}{4\Lambda b\kappa n}d\phi
\simeq
\frac{(-4\Lambda b\phi_0\kappa^3)^{n+2}}{16(n+2)\Lambda^2 b^2\kappa^4 n}\,,
\end{equation}
where $\phi_\text{i}$ is the value of the field at the end of inflation, and we have made use of the fact that the Ricci scalar is nearly constant during inflation and changes slower than the field. Thus,
\begin{equation}
(1-n_s)=\frac{2(n+1)}{(n+2)\mathcal N}\,,\quad
r=\frac{8\phi_0^2\kappa^2}{\mathcal N^2(n+2)^2}\,.
\end{equation}
In order to satisfy the data from Planck, $n$ must be large. In the limit of $1\ll n$ one has
\begin{equation}
(1-n_s)\simeq\frac{2}{\mathcal N}\,,\quad
r=\frac{8\phi_0^2\kappa^2}{\mathcal N^2 n^2}
\,,\quad 1\ll n\,,
\end{equation}
and since $\phi_0^2\kappa^2$ is on the order of the unit, also the tensor-to-scalar ratio is enough small.

\section{Mimetic gravity}

Motivated by the recent interest in mimetic gravity, in the present section we aim to contextualize $f(R,\phi)$-gravity within such a framework. In Refs.~\cite{m1, m2}, an approach for expressing the metric in terms of new degrees of freedom which isolates the conformal degree of freedom in a covariant way has been proposed (see ~\cite{matarrese} for further discussions, particularly on the role of disformal transformations in mimetic gravity, and related frameworks). The physical metric which describes the gravitational systems reads
\begin{equation}
g_{\mu\nu}=-\tilde g_{\mu\nu}\tilde g^{\alpha\beta}\partial_\alpha\varphi\partial_\beta\varphi\,,
\label{metricphi}
\end{equation}
where $\tilde g_{\mu\nu}$ is an auxiliary metric and $\varphi$ is a scalar field introduced by means of its first derivative. In this way, the metric and therefore the action of the theory are invariant under conformal transformations of $\tilde g_{\mu\nu}$ which take the form $\tilde g'_{\mu\nu}=\Omega(t, {\bf x})^2g_{\mu\nu}$, $\Omega(t, {\bf x})$ being a generic function of the coordinates.
As a direct consequence of (\ref{metricphi}) one has
\begin{equation}
g^{\mu\nu}\partial_\mu\varphi\partial_\nu\varphi=-1\,.\label{phiconst}
\end{equation} 
Such a property can be imposed on the theory by adding a Lagrange multiplier term to the action.
Let us consider mimetic $f(R,\phi)$-gravity in (\ref{main}), with the general action being given in the following form (see also ~\cite{OdMim, dmim} for further work on the subject)
\begin{equation}
I=\int_\mathcal{M}\left[\frac{f(R,\phi)}{2}-\frac{\omega(\phi)\partial^{\mu}\phi\partial_\mu\phi}{2}-V(\phi)\right]\sqrt{-g\left(\tilde g_{\mu\nu},\varphi\right)}dx^4\,.
\end{equation}
In the above, $\mathcal M$ is the space-time manifold, while the metric $g_{\mu\nu}=g_{\mu\nu}(\tilde g_{\mu\nu},\varphi)$ and its determinant $g\equiv g(\tilde g_{\mu\nu},\varphi)$ are functions of the auxiliary metric $\tilde g_{\mu\nu}$ and the field $\varphi$. It should be noted that the auxiliary metric never appears explicitly. Here, it is understood that the Ricci scalar is defined with respect to the physical metric $g_{\mu\nu}$ and therefore is also a function of of the auxiliary metric and the scalar field $\varphi$. \\

Varying with respect to $\tilde g_{\mu\nu}$ and $\varphi$ yields
\begin{equation}
G_{\mu\nu}=\frac{1}{F(R,\phi)}\left(T_{\mu\nu}^{\phi}+T_{\mu\nu}^{\text{MG}}+\tilde T_{\mu\nu}\right)\,,\label{fieldeq}
\end{equation}
where 
$G_{\mu\nu}$ is the Einstein's tensor, given by $G_{\mu\nu}=R_{\mu\nu}-g_{\mu\nu}R/2$ with $R_{\mu\nu}$ the Riemann tensor, and
$F(R,\phi)$ is given by (\ref{Fprime}). $T^{\phi}_{\mu\nu}$ is the stress energy tensor of the scalar field $\phi$ and modifications to Einstein's gravity are encoded in the tensor $T^{\text{MG}}_{\mu\nu}$
\begin{equation}
T^{\phi}_{\mu\nu}=\omega(\phi)\left[\partial_{\mu}\phi\partial_{\nu}\phi-\frac{1}{2}g_{\mu\nu}\partial^\alpha\phi\partial_{\alpha}\phi\right]-g_{\mu\nu}V(\phi)\,,
\end{equation}
\begin{equation}
T^{\text{MG}}_{\mu\nu}=\left(\frac{g_{\mu\nu}}{2}(f(R,\phi)-F(R,\phi)R)+\nabla_\mu\nabla_\nu F(R,\phi)-g_{\mu\nu}\Box F(R,\phi)\right)\,.
\end{equation}
Here,
$\nabla_\mu$ and $\Box$ are the covariant derivative and the d'Alambertian associated with the metric $g_{\mu\nu}\equiv g_{\mu\nu}(\tilde g_{\mu\nu},\varphi)$.
Furthermore, the tensor $\tilde T_{\mu\nu}$ reads
\begin{equation}
\tilde T_{\mu\nu}=-\left(F(R,\phi) G-T^{\phi}- T^{\text{MG}}\right)\partial_\mu\varphi\partial_\nu\varphi\,,\label{tildeT}
\end{equation}
where $G\,,T^\phi$ and $T^{\text{MG}}$ are the traces of the Einstein's tensor, the stress energy tensor of $\phi$ and of $T^{\text{MG}}_{\mu\nu}$, respectively,
\begin{eqnarray}
G&=&-R\,,
\nonumber\\
T^{\phi}&=&
-\omega(\phi)
\partial^{\mu}\phi\partial_{\mu}\phi-4V(\phi)\,,
\nonumber\\
T^{\text{MG}}&=&\frac{1}{8\pi G_N}\left[2\left(f(R,\phi)-F(R,\phi)R\right)-3\Box F(R,\phi)\right]\,.
\end{eqnarray}
By taking the covariant derivative of (\ref{fieldeq}), and recalling that $\nabla^\mu G_{\mu\nu}=0$ and  $\nabla^{\mu}T_{\mu\nu}^{\phi}=0$, we further obtain that
\begin{equation}
\nabla^\mu\left[\left(F(R,\phi)G-T^{\phi}-T^{\text{MG}}\right)\partial_{\mu}\varphi\right]
\equiv
\frac{1}{\sqrt{-g}}\partial_\kappa\left[\sqrt{-g}\left(F(R,\phi) G-T^{\phi}- T^{\text{MG}}\right)g^{\kappa\sigma}\partial_{\sigma}\varphi\right]
=0\,.\label{phieq}
\end{equation}
Note that the trace of the field equations (\ref{fieldeq}) leads to
\begin{equation}
(F(R,\phi)G-T^{\phi}-T^{\text{MG}})(1+g^{\mu\nu}\partial_\mu\varphi\partial_\nu\varphi)=0\,,
\end{equation}
which is automatically satisfied when (\ref{phiconst}) holds true, even when $(F(R,\phi)G-T^{\phi}- T^{\text{MG}})\neq 0$.

Here, we stress that it must be $0<F(R,\phi)$ to have a positive defined effective Newton constant in $\kappa^2\equiv 8\pi G_N$, such that $G_N^{\text{eff}}=G_N/F(R,\phi)$. In this respect, the $f(R,\phi)$-models (\ref{model1})
 and (\ref{model2}) avoid the ``antigravity'' in the corresponding $f(R)$-models (\ref{main}) and (\ref{HS})
at small curvatures after the end of inflation\footnote{Such a problem is not present in the original formulation of these models for the dark energy issue, where the history of the universe belongs to $R_0<R$.} when $\phi\rightarrow 0^{-}$ and $F(R,\phi)\simeq 1/\kappa^2$.\\

In the context of mimetic gravity, being the field $\varphi$ not fixed \textit{a priori}, one is faced with a wider class of solutions, as opposed to the simpler case where $F(R,\phi)G _{\mu \nu} = (T _{\mu \nu} ^{\phi} + T _{\mu \nu} ^{\text{MG}})$. A particularly interesting case arises when one considers a Friedmann-Robertson-Walker metric (\ref{metric}), which combined with (\ref{phiconst}) yields:
\begin{eqnarray}
\varphi = t + t _0 \ ,
\end{eqnarray}
In the above, $t _0$ is an integration constant which can be safely set equal to 0. Hence, Eq.(IV.9) reads:
\begin{eqnarray}
-\partial _t\left (a ^3\left (F(R,\phi)G - T ^{\phi} - T ^{\text{MG}} \right ) \right ) = 0 \ .
\end{eqnarray}
From the above we obtain:
\begin{eqnarray}
\left (F(R,\phi)G - T ^{\phi} - T ^{\text{MG}} \right ) = -\frac{c _0}{a ^3} \ ,\label{c0}
\end{eqnarray}
where $c _0$ is an integration constant. $\tilde{T} _{\mu \nu}$ can then be viewed as the stress-energy tensor for an additional component: 
\begin{equation}
\tilde{T} _{\mu \nu} = (\rho + p)u _{\mu}u _{\nu} + pg _{\mu \nu}\,. 
\end{equation}
The energy density, pressure and four-velocity of this new component are given respectively by:
\begin{eqnarray}
\rho = -\left (F(R,\phi)G - T ^{\phi} - T ^{\text{MG}} \right ) \ \ \ , \ \ \ p = 0 \ \ \ , \ \ \ u _{\mu} = \partial _{\mu}\varphi .
\end{eqnarray}
One can then see that $\tilde{T} _{\mu \nu}$ effectively describes a pressureless component, whose energy density decreases as $1/a ^3$. The additional degree of freedom provided by the scalar field $\varphi$, which encodes the conformal mode of gravity, can thus mimic cold dark matter. It is dynamical even in the absence of matter (i.e. ${\cal L} _m = 0$, as in the case we considered). The amount of dark matter is set by the integration constant $c_0$. \\

The mimetic gravity framework offers an alternative approach to solving some of the outstanding problems in modern cosmology. On cosmological scales, mimetic dark matter behaves precisely as collisionless cold dark matter, and as such is affected by gravitational instability ~\cite{m1}. On the other hand, it is known that the collisionless cold dark matter paradigm suffers from a number of shortcomings on small scales: the core-cusp problem ~\cite{corecusp}, the missing satellite problem ~\cite{missingsatellite}, and the ``too-big-to-fail" problem ~\cite{toobigtofail}, just to mention a few (see e.g. ~\cite{annika} for a recent review on the subject and more references). While in the particle dark matter framework these issues might be addressed by positing that dark matter is self-interacting and collisional (see e.g. ~\cite{selfinteracting} and references therein for further discussions), in the mimetic dark matter framework a possible solution is instead the addition of higher derivative (HD) terms for the scalar field $\varphi$ to the action. The HD terms (which are encoding UV physics) provide mimetic dark matter with a non-vanishing sound speed, and are effectively dissipation terms\footnote{Dark matter with significant dissipation has been studied recently in contexts other than mimetic gravity, both in particle ~\cite{dissipative,mirror} and non-particle frameworks ~\cite{komatsu} respectively.} ~\cite{ramazanov,imperfect,mirzagholi}. In fact, these HD terms might be crucial to avoid caustic singularities, from which the original mimetic dark matter framework suffers ~\cite{ramazanov}. More importantly, they have the effect of suppressing power on small scales, which has the potential to address the shortcomings of the collisionless cold dark matter framework on galactic and subgalactic scales ~\cite{ramazanov,mirzagholi}. \\

Another important observation is that mimetic gravity is equivalent to a class of Lorentz-violating generally covariant extensions of Einstein's General Relativity, known as Einstein-aether theories (EA hereafter, see ~\cite{szekeres} for the original formulation of the theory and ~\cite{aether} for more recent extensive discussions on this class of theories). In EA, Lorentz invariance is broken by a dynamical unit timelike vector $u ^{\mu}$ (the ``\textit{aether}"), which fixes a preferred rest frame at each spacetime point. In particular, mimetic gravity corresponds to the scalar formulation of the EA theory ~\cite{scalar1,scalar2} (see ~\cite{commentonscalar} for further discussions), where the aether vector is identified with the gradient of a scalar function, $u _{\mu} = \partial _{\mu}\varphi$. This scalar function corresponds to the scalar field encoding the conformal mode in mimetic gravity. The most general action for the scalar field, through the inclusion of the aforementioned HD terms, is constructed in ~\cite{scalar1}. Recently it was also noted that mimetic gravity can be identified and incorporated into the framework of covariant renormalizable gravity \cite{crgv}. \\

Another small-scale open question in mimetic gravity is whether this framework is able to account for the inferred flat rotation curves of spiral galaxies ~\cite{rubin}. The symmetries of the theory allow for the addition of a non-minimal coupling between matter and gravity, in the form of a coupling between the aether vector and a matter hydrodynamic flux. In the Newtonian limit such a term yields the phenomenology of MOND (see e.g. ~\cite{mond} for comprehensive reviews), and hence reproduces flat rotation curves and the Tully-Fisher relation ~\cite{tully}, among others. The phenomenology and constraints on such a coupling remain to be explored. \\

The mimetic gravity scenario can be successfully integrated with $f(R)$ gravity, and also with $f(R,\phi)$. Provided that the $f(R,\phi)$ (non-mimetic) theory is ghost-free, the corresponding mimetic formulation should also presumably be ghost-free, since the addition of a Lagrange multiplier term to the action is not expected to spoil such property (although it should be remarked that such a statement remains to be checked) ~\cite{OdMim}. One can then see how $f(R,\phi)$-gravity can successfully be incorporated in the mimetic gravity framework to additionally introduce a dark matter component to the theory, a crucial element in our current understanding of cosmology. Furthermore, it is possible to extend this picture by including a potential for the mimetic scalar field $\varphi$. Given that in an Friedmann-Robertson-Walker universe $\varphi$ can be identified with time, such an addition effectively introduces a time-dependent energy density, allowing one to realize any given evolutionary history of the Universe. Any potential $V(\varphi)$ can be used to reconstruct a function $f(R,\phi)$ which gives the corresponding evolution, as shown in ~\cite{OdMim}. In particular, by suitably choosing the form of the potential it is possible to construct an unified and consistent description of inflation with graceful exit, the current epoch of acceleration presumably sourced by dark energy, and a bouncing non-singular universe. \\

In the context of mimetic gravity, our inflationary models (\ref{model1}) and (\ref{model2}) acquire a dark matter term when $\tilde T_{\mu\nu}\neq 0$,
\begin{equation}
H\simeq\sqrt{\frac{\Lambda}{3}+\frac{c_0\kappa^2}{3a^3}}\,,
\end{equation}
where $c_0$ is fixed by (\ref{c0}).
The problem with this formulation is that, since the dark matter at the beginning of inflation is bounded above by the Planck mass, following inflation its contribution will be completely shifted away from the cosmological scenario. Various possible solutions can be envisioned, and here we will briefly discuss two possible ones. \\

In a first solution, note that when $\tilde T_{\mu\nu}=0$, $f(R,\phi)$-mimetic gravity leads to Eqs. (\ref{EOM1})--(\ref{EOM2}) and we recover the same results illustrated in the preceding section. Inflation ends when $\phi\rightarrow 0^-$ and the models (\ref{model1}) and (\ref{model2}) feature a ``phase transition'' from a high curvature regime ($f(R,\phi)\simeq R-2\Lambda$) to a low curvature one ($f(R,\phi)\simeq R$). One may assume that while inflation is realized at $\tilde T_{\mu\nu}=0$, the Friedmann universe belongs to the sector $T_{\mu\nu}\neq 0$ of the theory, and dark matter emerges only at the end of inflation. Another possible way of protecting dark matter from decay during inflation is to couple the two scalar fields $\varphi$ and $\phi$ through a term of the form $\varphi F(\phi)$, as discussed in \cite{m1}. \\

Another open question concerns generating the observed radiation and baryonic content in the universe, including the observed baryon-antibaryon asymmetry. This can presumably be generated by gravitational particle production following the end of inflation, through direct coupling of other fields to the scalar field $\varphi$, or through fluctuation-dissipation dynamics inherent to the scalar fields $\phi$ and $\varphi$ (see \cite{bererauno} and \cite{bereradue} for further discussions on the topic and the implementation of a model of dissipative leptogenesis). We plan to explore these and other ideas in a forthcoming paper.

\section{Conclusions}

In the present paper, we have studied inflation in the context of $f(R,\phi)$-theories of gravity, where a scalar field is coupled to gravity. This class of theories is very interesting, given that one can use the $f(R)$-gravity sector to reproduce a variety of cosmological scenarios (in our specific case, accelerated cosmology at high curvatures and Einstein’s gravity at low curvatures), while a dynamical scalar field allows for one to move between one scenario and another. We note that the $f(R)$-formulations of the models under investigation (namely, exponential gravity and the so called Hu-Sawiki model with power-law corrections to Einstein's gravity) belong to a class of viable models for the dark energy epoch which the universe undergoes today, where the appearance of an effective cosmological constant easily supports an (eternal) accelerated de Sitter expansion. Within the same models (perhaps in the attempt to unify the inflationary scenario with the dark energy epoch), one may reproduce the “false vacuum” of inflation by an effective cosmological constant, but a mechanism to make inflation unstable is necessary. In this respect, the introduction of a dynamical field induces a “phase transition” in the models and inflation ends when the effective cosmological constant disappears. \\

We have explicitly calculated the spectral indices and tensor-to-scalar ratio in the given models, starting from their first principle formulation in these kind of theories. We find that for the theory to give the correct amount of inflation (namely, a number of $e$-folds sufficiently large to allow for thermalization of the entire observable universe) and at the same time generate a spectral index in agreement with cosmological data, the magnitude of the tensor-to-scalar ratio (which is quadratic in one of the slow-roll parameter) is particularly small  (note that the same occurs in pure modified gravity but not in scalar field inflation within Einstein's framework) but non-vanishing: this occurs by virtue of the fact that the energy scale of inflation is sub-Planckian, while the magnitude of the scalar field can exceed the Planck scale. For recent work on $f(R,\phi)$-inflation see also ~\cite{unor, tur, pa}. \\

The minimal formulation we have considered does not contain a dark matter candidate. To address this point, we have then considered extensions of such a model within the mimetic gravity framework, where dark matter appears as an integration constant of the equations of motion. In the minimal mimetic gravity formulation, mimetic dark matter behaves precisely as collisionless cold dark matter. In the light of the issues which collisionless cold dark matter faces on small scales, we have discussed possible extensions of the mimetic gravity framework which allow to deal with these shortcomings, and at the same time explain a number of observations, the origin of which is usually attributed to particle dark matter (for instance, the inferred flat rotation curves and the Tully-Fisher relation). The extensions we have discussed were theoretically motivated by the equivalence between the original formulation of mimetic gravity and the Einstein-aether class of Lorentz-violating theories of gravity. We have further expounded how the extension of $f(R,\phi)$ inflation within the non-minimal mimetic gravity framework allows one to realize basically any evolutionary history of the Universe. Finally, we have commented on possible ways of protecting dark matter from decay during inflation, and generating the observed baryonic and radiation content of the Universe. \\

To conclude, the model we have explored introduces two additional scalar degrees of freedom to the framework of $f(R)$ gravity. A first one allows one to move between two different cosmological scenarios (accelerated expansion and Einstein gravity at high and low curvature respectively), while the second one endows the model with a natural dark matter candidate (which can address some of the small-scale tensions with collisionless cold dark matter), and can be used to reproduce any desired background cosmological expansion. While introducing extra degrees of freedom might seem a high price to pay, we have shown that if used appropriately such degrees of freedom allow a more natural implementation of a unified expansion history, while at the same time providing a candidate for the missing dark matter in the Universe.

\section*{Acknowledgements}

We would like to thank Sergei Odintsov and Sergio Zerbini for comments and suggestions.
S.V. would like to thank Amel Durakovi\'{c} for useful discussions, and the hospitality of the Niels Bohr International Academy and in particular of Poul Henrik Damgaard while this work was being completed.

\end{document}